\documentclass[preprint]{revtex4}

\usepackage{graphicx}% Include figure files
\usepackage{dcolumn}% Align table columns on decimal point
\usepackage{bm}% bold math
\usepackage{color}

\begin{document}

\title{Collective Molecular Dynamics in Proteins and Membranes}

\author{Maikel C.~Rheinst\"adter}\email{RheinstadterM@missouri.edu}

\affiliation{Department of Physics and Astronomy, University of
Missouri-Columbia, Columbia, MO 65211, U.S.A.}

\date{\today}% It is always \today, today,
             %  but any date may be explicitly specified

%------------------------------------------------------------------------------------
\begin{abstract}
The understanding of dynamics and functioning of biological
membranes and in particular of membrane embedded proteins is one of
the most fundamental problems and challenges in modern biology and
biophysics.  In particular the impact of membrane composition and
properties and of structure and dynamics of the surrounding
hydration water on protein function is an upcoming hot topic, which
can be addressed by modern experimental and computational
techniques. Correlated molecular motions might play a crucial role
for the understanding of, for instance, transport processes and
elastic properties, and might be relevant for protein function.
Experimentally that involves determining dispersion relations for
the different molecular components, i.e., the length scale dependent
excitation frequencies and relaxation rates.  Only very few
experimental techniques can access dynamical properties in
biological materials on the nanometer scale, and resolve dynamics of
lipid molecules, hydration water molecules and proteins and the
interaction between them.  In this context, inelastic neutron
scattering turned out to be a very powerful tool to study dynamics
and interactions in biomolecular materials up to relevant nanosecond
time scales and down to the nanometer length scale. We review and
discuss inelastic neutron scattering experiments to study membrane
elasticity and protein-protein interactions of membrane embedded
proteins.

%Scattering techniques are a powerful too to determine dynamical and
%structural properties of proteins and membranes. The combination of
%various techniques enlarges the window to length scales from the
%nearest-neighbor distances of lipid molecules to more than
%micrometers, covering time scales from picoseconds to seconds, also
%covering intra- and inter-protein dynamics of membrane embedded
%proteins. The main research objective is to quantify collective
%molecular fluctuations in these systems and to establish
%relationships to properties and functional aspects, and eventually
%biological functions of the bilayers. The experiments allow
%contributing to one of the most fundamental questions in modern
%biophysics and biology, i.e., to study dynamics of membrane embedded
%proteins and how bilayer composition and properties affect protein
%function.
\end{abstract}

\maketitle

%------------------------------------------------------------------------------------
\section{Introduction}
\begin{figure}[p]
\centering
\resizebox{1.00\columnwidth}{!}{\rotatebox{0}{\includegraphics{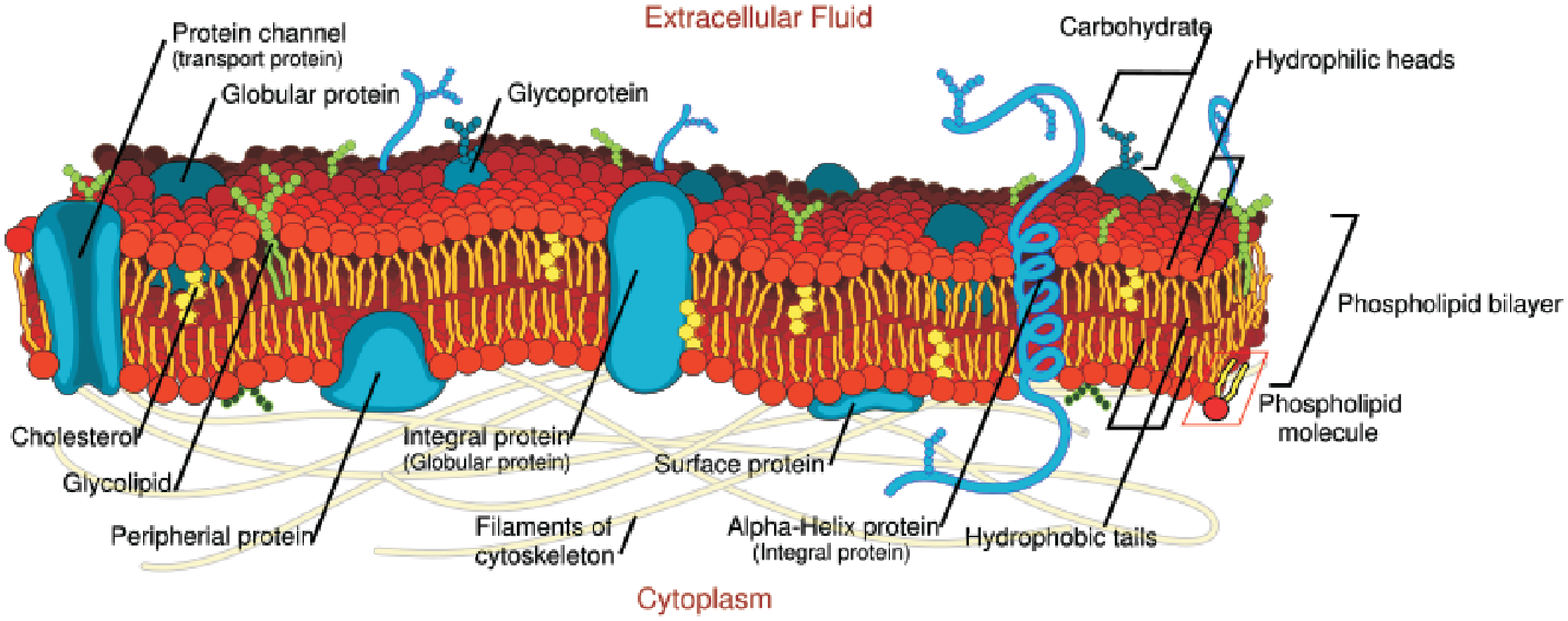}}}
\caption[]{} \label{cellmembrane}
\end{figure}

Dynamics in biological membranes occur on a wide range of length and
time scales and involves interactions between the different
constituents, such as lipids, cholesterol, peptides and proteins, as
sketched in Figure~\ref{cellmembrane}. Often, the structure of these
systems is relatively well known and the corresponding experimental
techniques, such as x-ray crystallography, Nuclear Magnetic
Resonance (NMR), and also Atomic Force Microscopy (AFM) have
developed standard techniques in many disciplines. The access to
dynamical properties turned out to be more difficult. Dynamical
properties are often less well understood in biomolecular systems,
but are important for many fundamental biomaterial properties such
as, e.g., elasticity properties and interaction forces, and might
determine or strongly affect certain functional aspects, such as
diffusion and parallel and perpendicular transport through a
bilayer, and also protein function. New developments and
improvements in neutron scattering instrumentation, sample
preparation and environments and, eventually, the more and more
powerful neutron sources open up the possibility to study collective
molecular motions on lateral length scales between micrometers down
to a few Angstroems (\AA). The fluctuations are quantified by
measuring the corresponding dispersion relations, i.e., the wave
vector-dependence of the excitation frequencies or relaxation rates.
Because biological materials lack an overall crystal structure, in
order to fully characterize the fluctuations and to compare
experimental results with membrane theories, the measurements must
cover a very large range of length and time scales
\cite{RheinstaedterJAV:2006}. The combination of various inelastic
scattering techniques, such as inelastic neutron and x-ray
scattering, and dynamic light scattering, enlarges the window of
accessible momentum and energy transfers - or better: accessible
length and time scales - and allows one to study structure and
dynamics from the nearest-neighbor distances of lipid molecules to
more than 100 nanometers (three orders of magnitude), covering time
scales from about one tenth of a picoseconds to almost 1~second
(twelve orders of magnitude). The spectrum of fluctuations in model
membranes for example covers the long wavelength undulation and
bending modes of the bilayer with typical relaxation times of
nanoseconds and lateral length scales of several hundred lipid
molecules, down to the short-wavelength, picosecond density
fluctuations involving neighboring lipid molecules.

Even that membranes are now studied extensively for more than three
decades \cite{Lipowsky:1995}, the field is currently boosted by the
inter disciplinary interest in life sciences. New and more powerful
techniques might give a quantitative access to molecular properties
of the bilayers. Often, phospholipid bilayers are used as model
systems to study fundamental properties
\cite{Salditt:2000,Krueger:2001,Salditt:2005}. But the trend
definitely goes to more complex and relevant systems, decorated with
cholesterol, peptides and proteins to eventually develop a better
understanding of biological membranes. Different model systems are
used in experiments and Molecular Dynamics Simulations (MD). In
scattering experiments, the weak signals arising from dynamical
modes can be multiplexed by using stacked membrane systems and large
system sizes, i.e., membrane patches in the order of micrometers and
larger. The advantage is the large sampling rate, which makes very
small signals visible, and the large ensemble average, which
averages over defects. The 'muli-scale' character of biological
membranes, i.e., that relevant dynamics occur on a large range of
length and time scales
\cite{Frauenfelder:1991,Fenimore:2004,Bayerl:2000}, poses particular
problems because different techniques and probes must be applied to
access the different lengths and frequencies. Scattering techniques
are particularly powerful in spatially ordered systems and have been
very successfully applied to study structure and dynamics (phonons)
in crystals. In soft-matter and biology, a periodic structure is
often missing or not well developed and new concepts and approaches
are required to study structural and dynamical properties. In
contrast to computer simulations, where different molecular
components and dynamics can well be extracted and distinguished, in
experiments there usually is a superposition of signals of different
molecular components and interactions, and also single-molecule and
collective dynamics. By using different sample geometries, these
signals might be separated by their position in reciprocal and
frequency space. Selective deuteration is used to distinguish
between incoherent and coherent dynamics. Because of more and more
powerful spectrometers, smaller and smaller system dimensions and
less and less material can be used in experiments. But experiments
are still orders of magnitude away from single molecule spectroscopy
of, e.g., membrane embedded proteins, which might be envisioned in
the next decade.

Even that computer power and access to powerful clusters has
drastically increased in the last couple of years, MD simulations
are still limited when it comes to large system sizes. Standard
membrane systems are single bilayers, and contain several hundred
lipid molecules, leading to patch sizes of about 100~\AA, and
equilibrated for about 100 nanoseconds \cite{Tarek:2001,Hub:2007}.
The system size, D, produces a cut off wavelength ($\lambda=2\pi/D$)
for the dynamical modes that can be observed and quantified. To
properly determine an excitation frequency or relaxation rate from
the time evolution of the scattering functions, the simulations must
cover at least two orders of magnitude more in maximal accessible
time, thereby limiting the accessible frequencies to about 1~ns. So
simulations are still very much limited to fast short wavelength
dynamics. When (single) proteins are simulated, the small system
size prevents to study the influence of bilayer dynamics to protein
dynamics and function, and also a possible protein-protein
interaction, which might be relevant for biological function. %To
%compare to experiments, system dimension must become larger by at
%least an order of magnitude.
Analytical theory for membranes is usually based on the theory
developed for liquid crystals
\cite{deGennes:1974,Kats:1993,Ribotta:1974}. It is challenging how
microscopic properties, such as the lipid composition of a bilayer,
affects macroscopic behavior. However, theories are very important
to determine properties from fitting experimental and computational
data.

Collective molecular motions impact, e.g., on properties and
functionalities of artificial and biological membranes, such as
elasticity, transport processes, and inter-protein interaction. Two
examples will be discussed in detail in the following:
\begin{enumerate}
\item Mesoscopic shape fluctuations in aligned multi lamellar stacks of
DMPC (1,2-dimyristoyl-sn-glycero-3-phoshatidylcholine) bilayers were
studied using the neutron spin-echo technique. From the dispersion
relation in the fluid phase, values for the bilayer bending rigidity
$\kappa$, the compressional modulus of the stacks $B$, and the
effective sliding viscosity $\eta_3$ could be determined. This
technique offers a novel approach to quantify the elasticity
parameters in membranes by direct measurement of dynamical
properties and also the impact of collective molecular motions on
membrane properties \cite{RheinstaedterPRL:2006}.

\item Very recently, interprotein motions in a carboxymyoglobin protein
crystal were reported from a molecular dynamics simulation [Phys.
Rev. Lett. 100, 138102 (2008)]. Experimental evidence for a
cooperative long range protein-protein interaction in purple
membrane (PM) was found by inelastic neutron scattering. The
dynamics was quantified by measuring the spectrum of the acoustic
phonons in the 2d Bacteriorhodopsin (BR) protein lattice. The data
were compared to an analytical model and the effective spring
constant for the interaction between protein trimers was determined
to be $k=53.49$ N/m. The experimental results are in very good
agreement to the computer simulations, which reported an interaction
energy of 1~meV \cite{RheinstadterPRL:2008}. In this case, inelastic
neutron scattering was used to study interactions between
constituents of a biological membrane.
\end{enumerate}

Transport through membranes is one of the most fundamental
functionalities of the bilayers
\cite{Huster:1997,Deamer:1986,Paula:1996,Lahajnar:1995}. While
transport channels or transport proteins are responsible for the
transport of larger molecules, small molecules, such as water, can
pass the bilayer with a certain probability. Despite the large body
of experimental and theoretical work, it turns out to be very
difficult to model the permeation of small molecules through
bilayers on a microscopic level because of the large number of
parameters involved. The solubility-diffusion model and the
transient pore model are possible mechanisms for permeation of small
molecules \cite{Paula:1996}: Pores seem to be the dominant
permeation mechanism for ions for thin membranes. Ion permeation by
partitioning and diffusion seems to become of greater importance as
membrane thickness increases. Neutral molecules, such as water,
cross by the solubility-diffusion mechanism because of their high
solubility in the hydrocarbon phase. Recently Nagle {\em et al.}
\cite{Nagle:2008} and Mathai {\em et al.} \cite{Mathai:2008}
presented in a combined study the theoretical and experimental
framework of a a three layer theory for the passive permeability
through bilayers. They link permeability to structural properties
and strongly correlate it with the area per lipid, A, rather than
with other structural quantities such as the thickness. Decades ago
Tr\"auble \cite{Trauble:1971} (and later addressed by Marsh
\cite{Marsh:1974}) discussed the movement of molecules across
membranes in terms of thermal fluctuations in the hydrocarbon chains
of the membrane lipids. The thermal motion of the hydrocarbon chains
results in the formation of conformational isomers, so-called
kink-isomers of the hydrocarbon chains. "Kinks" may be pictured as
mobile structural defects which represent small, mobile free volumes
in the hydrocarbon phase of the membrane. The free volume needed for
kink creation and movement might be created by propagating in-plane
density fluctuations of the lipid acyl chains
\cite{RheinstaedterPRL:2004}. In a first inelastic neutron
scattering experiment, the corresponding dispersion relation of the
acyl tail dynamics in DMPC bilayers showed a drastic softening in
the presence of ethanol molecules, which are known to increase
bilayer permeability. Future experiments will elucidate the
influence of ethanol on the collective dynamics in phospholipid
bilayers.

%------------------------------------------------------------------------------------
\section{Collective vs. single-molecule dynamics}
The spectrum of fluctuations in biomimetic and biological membranes
covers a large range of time and length scales
\cite{Koenig:1992,Koenig:1994,Koenig:1995,Pfeiffer:1989,Pfeiffer:1993,Lindahl:2000,Lipowsky:1995,Bayerl:2000,Salditt:2000},
ranging from the long wavelength undulation and bending modes of the
bilayer with typical relaxation times of nanoseconds and lateral
length scales of several hundreds lipid molecules to the short
wavelength density fluctuations in the picosecond range on nearest
neighbor distances of lipid molecules. Local dynamics in lipid
bilayers, i.e., dynamics of individual lipid molecules as vibration,
rotation, libration (hindered rotation) and diffusion, have been
investigated by, e.g., incoherent neutron scattering
\cite{Koenig:1992,Koenig:1994,Koenig:1995,Pfeiffer:1989,Pfeiffer:1993}
and nuclear magnetic resonance \cite{Nevzorov:1997,Bloom:1995} to
determine the short wavelength translational and rotational
diffusion constant. Collective undulation modes have been
investigated using neutron spin-echo spectrometers
\cite{Pfeiffer:1989,Pfeiffer:1993,Takeda:1999,RheinstaedterPRL:2006}
and dynamical light scattering
\cite{Hirn2:1999,Hirn:1999,Hildenbrand:2005}.

\begin{figure}[p]
\centering
\includegraphics[width=1.00\columnwidth,angle=0]{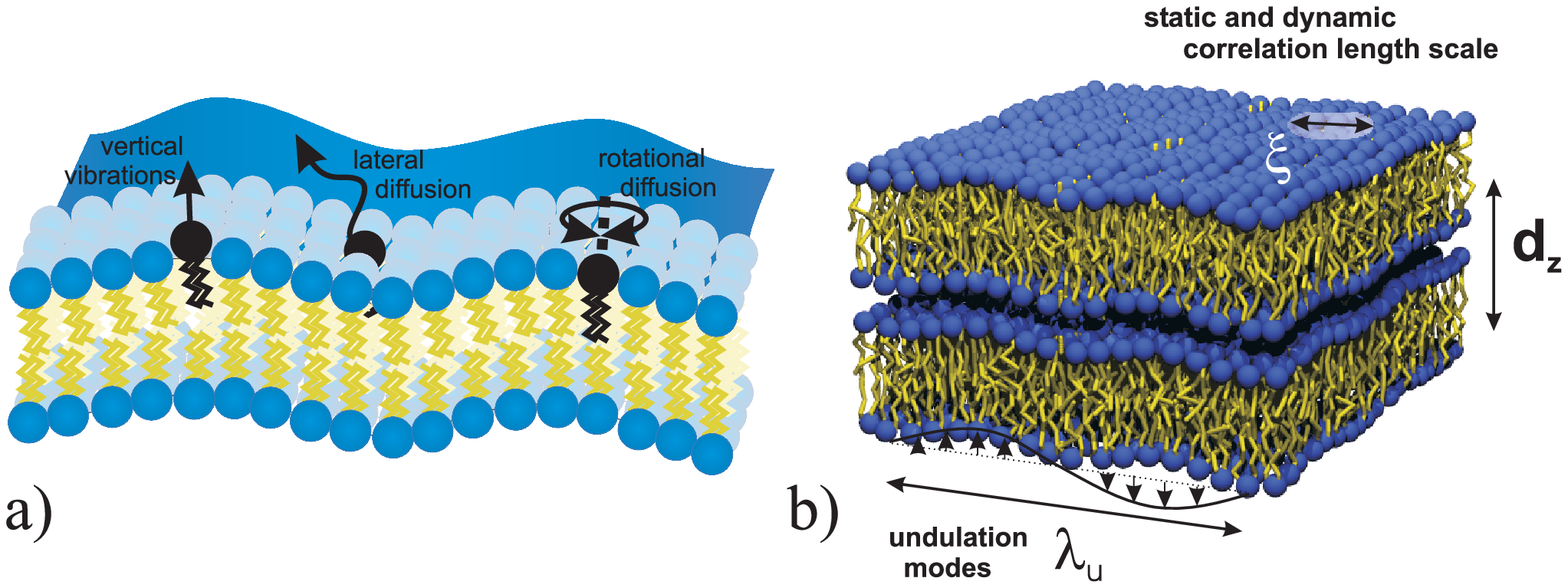}
\caption[]{} \label{membranedynamics.eps}
\end{figure}

Atomic and molecular motions in membranes can be classified as
local, autocorrelated, and collective, pair correlated dynamics.
Within the scattering formalism, autocorrelated dynamics is
described by the incoherent scattering function,
S$_{incoherent}(q,\omega)$, while the coherent scattering,
S$_{coherent}(q,\omega)$, describes dynamics involving different
molecules or atoms. Figure~\ref{membranedynamics.eps} exemplary
depicts some of the local and collective modes in a phospholipid
bilayer. Rotational and lateral diffusion, vibrations and rotations
of the single lipid molecules can be investigated by e.g.,
incoherent inelastic neutron scattering, nuclear magnetic resonance
or dielectric spectroscopy. On the other hand only coherent
inelastic neutron scattering or inelastic x-ray scattering are able
to elucidate the collective excitations such as, e.g., the short
wavelength density fluctuations or undulation modes of the bilayer.

\begin{figure}[p]
\centering
\includegraphics[width=1.00\columnwidth,angle=0]{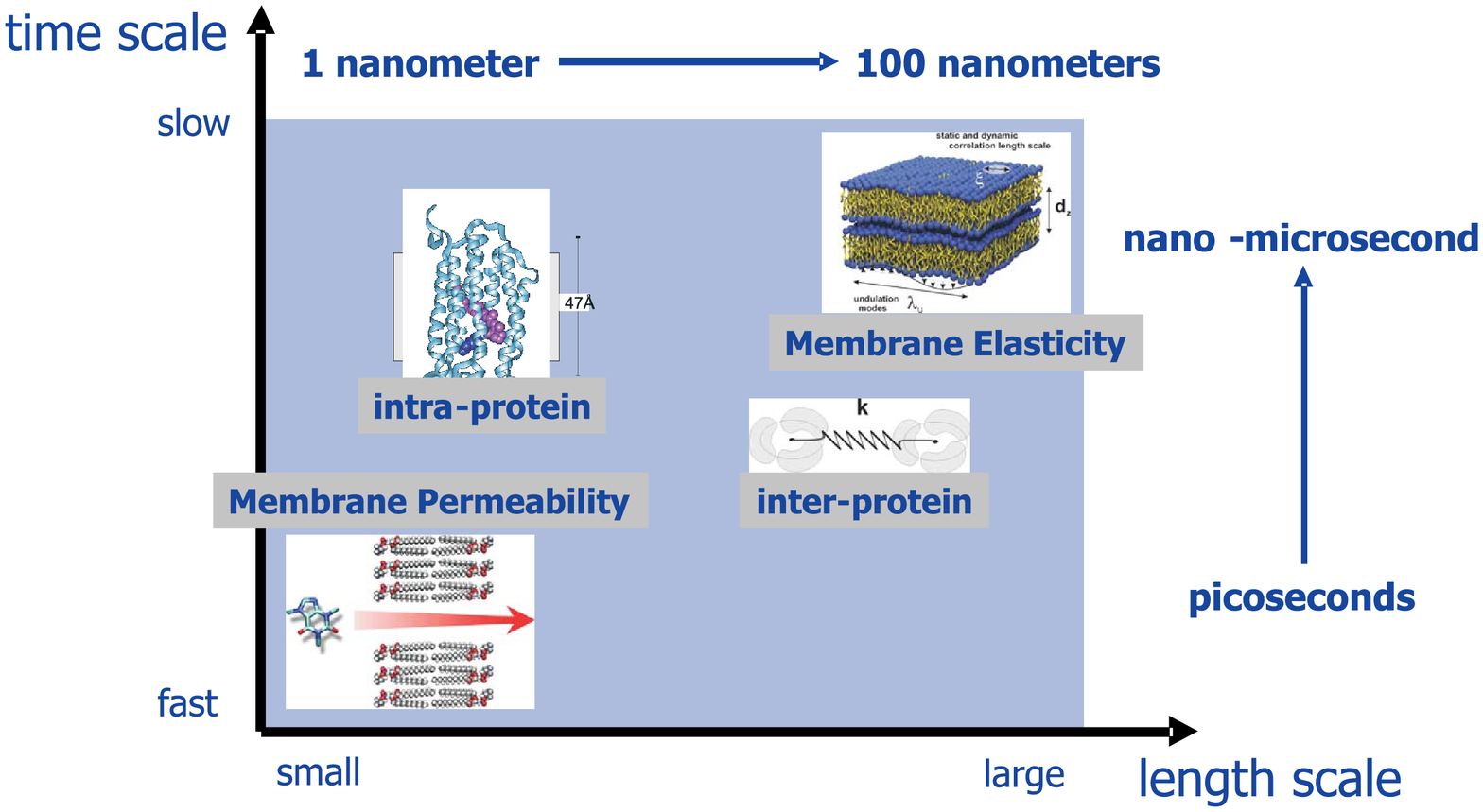}
\caption[]{} \label{Collective_Function_Graph}
\end{figure}

Experimentally, selective deuteration is used to emphasize the
incoherent, respective coherent scattering over other contributions
to the total sample scattering. While in protonated samples the
incoherent scattering is normally dominant and the
time-autocorrelation function of individual scatterers is accessible
in neutron scattering experiments, (partial) deuteration emphasizes
the coherent scattering and gives access to collective motions by
probing the pair correlation function. Computer simulations offer
direct access to incoherent and coherent properties by calculation
of S$_{incoherent}(q,\omega)$ and S$_{coherent}(q,\omega)$ from the
time evolution of atomic and molecular coordinates.
Figure~\ref{Collective_Function_Graph} visualizes some of the length
and time scales involved for collective dynamics and the
corresponding functional aspects. While permeability of bilayers
occurs on distances of neighboring lipid molecules, elasticity might
involve hundreds of membrane molecules. Coherent intra-protein
dynamics will most likely be faster than inter-protein dynamics,
which involves larger distances and in most cases a lipid mediated,
elastic interaction.

%------------------------------------------------------------------------------------
\section{Experimental}
Only recently, the first inelastic scattering experiments in
phospholipid bilayers to determine collective motions of the lipid
acyl chains and in particular the short wavelength dispersion
relation have been performed using inelastic X-ray \cite{Chen:2001}
and neutron \cite{RheinstaedterPRL:2004} scattering techniques. Note
that only scattering experiments give wave vector resolved access to
dynamical properties, what is important to associate excitation
frequencies and relaxation times with specific molecular components
and motions. In the case of single membranes the scattering signal
is usually not sufficient for a quantitative study of the inelastic
scattering. To maximize the scattering signal, multi lamellar
samples composed of stacks of several thousands of lipid bilayers
separated by layers of water, resulting in a structure of smectic A
symmetry, were prepared. The high orientational order of the samples
which gives rise to pronounced Bragg peaks and excitations is a
prerequisite to a proper analysis of the corresponding correlation
functions. Highly oriented multi lamellar membrane stacks of several
thousands of lipid bilayers were prepared by spreading lipid
solution of typically 25~mg/ml lipid in trifluoroethylene/chloroform
(1:1) on 2'' silicon wafers. About twenty such wafers separated by
small air gaps were combined and aligned with respect to each other
to create a ''sandwich sample'' consisting of several thousands of
highly oriented lipid bilayers with a total mosaicity about
0.5$^{\circ}$), and a total mass of about 400~mg of deuterated DMPC.
The mosaicity of the sandwich is composed of the alignment of the
bilayers within the stack on one wafer and of the orientation of
different wafers with respect to each other. The use of well
oriented samples leads to well localized elastic and inelastic
signal in reciprocal space and allows to distinguish motions in the
plane of the membranes ($q_{||}$) and perpendicular to the bilayers
($q_z$). During the experiments, the membranes were kept  in a
"Humidity Chamber" to control temperature and humidity and hydrated
with D$_2$O from the vapor phase. The collective motions of the
lipid acyl chains have been emphasizes over other contributions to
the inelastic scattering signal by using partially, chain deuterated
lipids (DMPC -d54). The experiments that will be discussed here,
were conducted on two different types of neutron spectrometers,
namely triple-axis and spin-echo spectrometers. The accessible
length and time scales of these spectrometers are discussed in more
detail in Ref.~[\onlinecite{RheinstaedterJAV:2006}]. The present
paper focuses on the dynamics-property and dynamics-function aspects
of the experiments.

\subsection{Spin-echo Spectrometry}
Fluctuations on the mesoscopic scale are determined by the
elasticity parameters of the bilayers, i.e., the compressibility of
the stacked membranes, B, and the bending modulus $\kappa$. The
relaxations in this regime are in the nano-second time-range with
accompanying small $q$-vales. Spin-echo spectrometers turned out to
be highly suited for these experiments. The spin-echo technique
offers extremely high energy resolution from Larmor tagging the
neutrons. A neutron spin echo measurement is in essence a
measurement of neutron polarization \cite{Mezei:1980}. A polarized
neutron beam passes through a magnetic field perpendicular to the
neutron polarization. The neutron spin precesses before arriving at
the sample, acquiring a precession angle $\varphi_1$. At the sample,
the beam is scattered before passing through a second arm, acquiring
an additional precession angle $\varphi_2$ in the reversed sense.
For elastic scattering the total precession angle is
$\Delta\varphi=\varphi_1-\varphi_2=0$ for all incoming neutron
velocities. If the neutron scatters inelastically by a small energy
transfer $\hbar\omega$, there will be a linear change
$\Delta\varphi=\tau\cdot\omega$ with $\tau$ being a real time in the
case of quasielastic scattering. The spin-echo technique thus works
in the time domain and measures the intermediate scattering function
S($q_{||},t$) in contrast to the three-axis technique. For a
quasielastic response, assumed to have Lorentzian lineshape with
half-width $\Gamma$, the polarization will then show a single
exponential decay PNSE = P$_s e^{\Gamma t}$.

\subsection{Triple-axis spectrometry}
Fast motions in the ps time range due to sound propagation in the
plane of the bilayer are best measured on triple-axis spectrometers.
The energy of the incident and scattered neutrons is determined by
Bragg scattering from crystal monochromators, (graphite in most
cases). Advantages of triple-axis spectrometers are their relatively
simple design and operation and the efficient use of the incoming
neutron flux to the examination of particular points in $(q,\omega)$
space. By varying the three axes of the instrument, the axes of
rotation of the monochromator, the sample and the analyzer, the wave
vectors $k_i$ and $k_f$ and the energies E$_i$ and E$_f$ of the
incident and the scattered neutrons, respectively, can be
determined. The momentum transfer to the sample, and the energy
transfer, $\hbar\omega$, are then defined by the laws of momentum
and energy conservation to $q=k_f-k_i$ and
$\hbar\omega$=E$_i$-E$_f$. The accessible $(q,\omega)$ range is just
limited by the range of incident neutron energies offered by the
neutron guide as well as by mechanical restrictions of the
spectrometer.

%------------------------------------------------------------------------------------
\section{Elastic Properties}
According to linear smectic elasticity theory
\cite{Caille:1972,Lei:1995} thermal fluctuations in the fluid phase
of the membrane are governed by the free energy functional
(Hamiltonian) \cite{Caille:1972,Lei:1995,LeiThesis:1993}:
\begin{equation}
H = \int_A d^2r
\sum_{n=1}^{N-1}\left(\frac{1}{2}\frac{B}{d}(u_{n+1}-u_n)^2+\frac{1}{2}\kappa\left(\nabla^2_{||}u_n\right)^2\right)~,
\label{Hamiltonian}\end{equation} where $\kappa$ denotes the bilayer
bending rigidity, $A$ the area in the $xy$-plane, $N$ the number of
bilayers, and $u_n$ the deviation from the average position $n~d$ of
the $n$-th bilayer, $d$ is the lamellar spacing. $B$ and $K =
\kappa/d$ are elastic coefficients, governing the compressional and
bending modes of the smectic phase, respectively. A fundamental
length scale in these systems is given by the smectic penetration
length $\Lambda = \sqrt{K/B}$. Aligned lipid bilayers allow a
separate determination of both parameters $K$ and $B$
\cite{Lyatskaya:2001,Salditt:2003}.

\begin{figure}[p]
\centering
\resizebox{0.75\columnwidth}{!}{\rotatebox{0}{\includegraphics{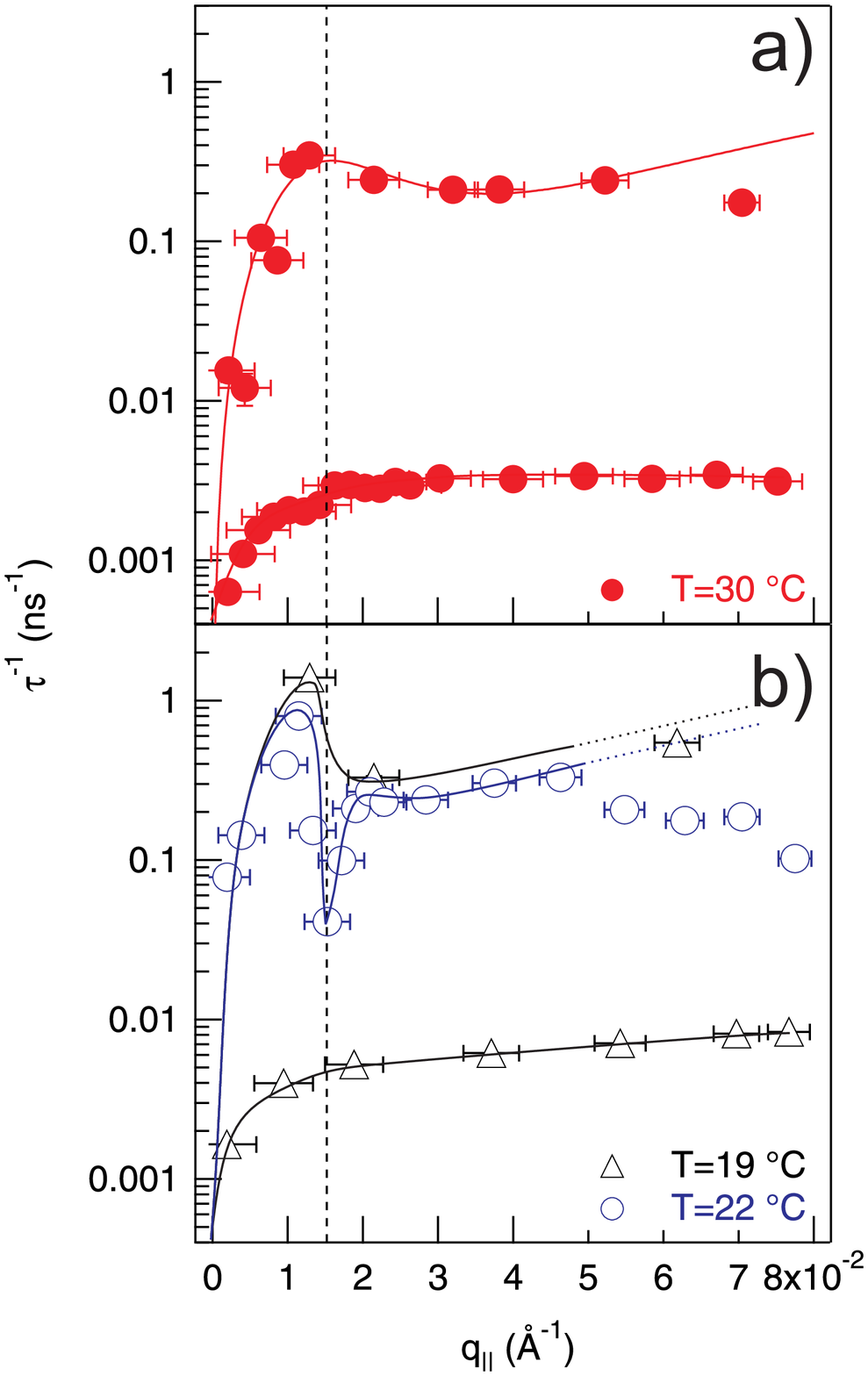}}}
\caption[]{} \label{dispersion_fluid_graph.eps}
\end{figure}

The spin-echo experiments were carried out at the IN11 and IN15
spectrometers, at the cold source of the high flux reactor of the
Institut Laue-Langevin (ILL) in Grenoble, France. Wavelength bands
centered at $\lambda$=7.4\AA\ and $\lambda$=14\AA\ with $\Delta
\lambda/ \lambda \simeq 0.15$ (FWHM), respectively, have been set by
a velocity selector. The intermediate scattering function
S(q$_{||}$,t) was measured for spin-echo times of 0.001ns$<$t$<$20ns
for IN11 and 0.01ns$<$t$<$200ns for IN15. Data have been taken at
three different temperatures, at 19$^{\circ}$C, in the gel (ripple,
P$_{\beta'}$) phase of the phospholipid bilayers, at 22$^{\circ}$C,
just above the temperature of the main transition in deuterated
DMPC-d54 (at T$_m\approx 21.5$$^{\circ}$C), and at 30$^{\circ}$C,
far in the fluid L$_{\alpha}$ phase of the membranes and above the
regime of so-called anomalous swelling. The corresponding lamellar
$d$ spacings were $d$=56~\AA, 60~\AA\ and 54~\AA\ (gel,
22$^{\circ}$C and fluid), respectively. Two relaxation processes,
one at about 10~ns ($\tau_1$) and a second, slower process at about
100~ns ($\tau_2$) were observed. The relaxation rates $\tau_1^{-1}$
and $\tau_2^{-1}$ in the gel and the fluid phase are depicted in
Fig.~\ref{dispersion_fluid_graph.eps} (a) and (b). Both relaxation
branches are dispersive. The fast process shows a q$_{||}^2$
increase at small q$_{||}$ values and a bend at about
q$_{||}\approx$0.015\AA$^{-1}$. The dispersion in the gel phase and
close to the phase transition in
Fig.~\ref{dispersion_fluid_graph.eps} (b) appear to be more
pronounced as compared to 30$^{\circ}$C dispersion. A soft mode
appeared in the T=22$^{\circ}$C dispersion, indicating a significant
softening of the bilayer at a well defined wave number. The slow
branches at T=19$^{\circ}$C and 30$^{\circ}$C also show increasing
relaxation rates with increasing q$_{||}$ values, but with a
distinct non-polynomial behavior.

The dispersion relation of the fast branch with relaxation rates
between 1 and 10~ns can be attributed to undulation dynamics.
Qualitatively, at very small $q_{||}$-values, the membranes behave
as liquid films and their dynamics is basically determined by the
viscosity of the water layer in between the stacked membranes ({\em
film regime}). With increasing $q_{||}$ there is a transition into a
{\em bulk-elasticity} regime where the dynamics depends on the
elastic properties of the lipid bilayers. At this point, the
dispersion bifurcates in two relaxation branches. The faster one,
which is out the experimentally accessible time window of the
spin-echo spectrometer, is mainly determined by the compressional
modules $B$. The slower one, which is observed here, can be assigned
to the bending modulus $\kappa$. The slow dispersion branch with
relaxation rates of about 100~ns could be attributed to a surface
relaxation mode \cite{Bary-Soroker:2006,Bary-Soroker:2007}, which is
particular to a stack of membranes.

Following the idea of Ribotta \cite{Ribotta:1974}, the relaxation
rates of the undulations can be described by:
\begin{equation}
\label{ribotta} \tau^{-1}(q_{||}) =\frac{\kappa/d}{\eta_3} q_{||}^2
\frac{q_{||}^4+(\pi/(\Lambda
D))^2}{q_{||}^4+\frac{1}{\mu\eta_3}(\pi/D)^2} %~.
\end{equation}
($\eta_3$ is the layer sliding viscosity), and following results
were  obtained: $\kappa=14.8\pm 8$k$_B$T, $\Lambda=10.3 \pm
2.3$~\AA, $\eta_3=0.016\pm 0.0006$~Pa~s. B was calculated to
B=1.08~10$^7$J/m$^3$ ($d$=54~\AA). These values agree quite well
with values reported in the literature
\cite{Ollinger:2005,Petrache:1998,Lindahl:2000,Pabst:2003}. The
resulting effective sliding viscosity of the membrane system
$\eta_3$ was found to be 16 times higher than that of water, what
points to different properties of the interstitial hydration water,
as compared to bulk water. Note that Eq.~\ref{ribotta} does not
describe a pure undulation mode, which is probed at q$_z$ values of
$q_z=2\pi/d$, only. If the scattering is probed at finite components
$\delta q_z:=(q_z-2\pi/d)$ or measured with a relaxed q$_z$
resolution, there is a mixing of {\em baroclinic} modes, which are
distinctly slower than a pure undulation because they involve a
relocation motion of the water layer. The parameter $D$ describes an
effective finite-size cutoff-length, which was related to the
instrumental resolution $D=\pi/\Delta q_z$.

From the soft-mode in the fluid dispersion, it seems that the well
known softening of phospholipid membrane upon approaching the main
phase transition temperature from the fluid phase, i.e., the regime
of ''critical swelling'' or ''anomalous swelling''
\cite{Pabst:2003}, occurs on a well defined length scale, only
($2\pi/q_{||}\approx 420$~\AA). Using atomic force microsopy (AFM),
pronounced ripples were observed in the P$_{\beta'}$ phase of
stacked membranes, with an increasing ripple periodicity $\Lambda_r$
when approaching the temperature of the main transition at
$T_m=24~^{\circ}C$ \cite{Schafer:2008}. Close to $T_m$, coexisting
metastable ripples
with $2\Lambda_r\approx 420$~\AA\ were observed. %The length scale of
%these ripples agrees well with the length scale of a soft mode in
%the dispersion relation of the long wavelength undulation modes
%reported in a recent inelastic neutron scattering study and might be
%responsible for the well known softening of phospholipid bilayers in
%the range of the critical swelling. The ripple period $\Lambda_r$
%increases also with increasing osmotic pressure, most likely due to
%an increasing interaction between the bilayers in the stack.
The experiments therefore point to coexisting {\em nanodomains} with
sizes of less than $50~nm$ in the range of {\em critical swelling}
\cite{Chen:1997,Nagle:1998,Mason:2001} of phospholipid bilayers. The
existence of coexisting small gel and fluid domains has also been
argued by preceding AFM investigations \cite{Xie:2002,Tokumasu} to
compensate the large stress which occurs at $T_m$ due to the volume
difference of the two phases. The softening in the range of the
phase transition is most likely coupled to the occurrence of these
metastable $2\Lambda_r$ ripples. Bending of the bilayers might occur
mainly in the interfaces between two metastable ripples where the
bending modulus can be expected to be softer because structure and
interactions are likely to be much less well defined in the
interface. Not much energy would be needed to slightly change the
tilt angle between to ripple flanks. Note that the relation between
critical swelling and softening of the bilayers, and the formation
of a low temperature ripple phase is not trivial
\cite{Mason:2001,Pabst:2004} and the anomalous swelling is not
directly coupled to the formation of ripples. Inelastic experiments
in different systems are definitely needed to better understand the
impact of the soft-mode in the DMPC dispersion curve. The most
important contribution of the NSE technique is that the 'stiffness'
of the bilayers is determined as a function of internal length
scale. It seems that there is no global softening, but softening
occurs on of certain length scales, only.

%------------------------------------------------------------------------------------
\section{Collective Protein Interaction}
\begin{figure}[p] \centering
\resizebox{1.00\columnwidth}{!}{\rotatebox{0}{\includegraphics{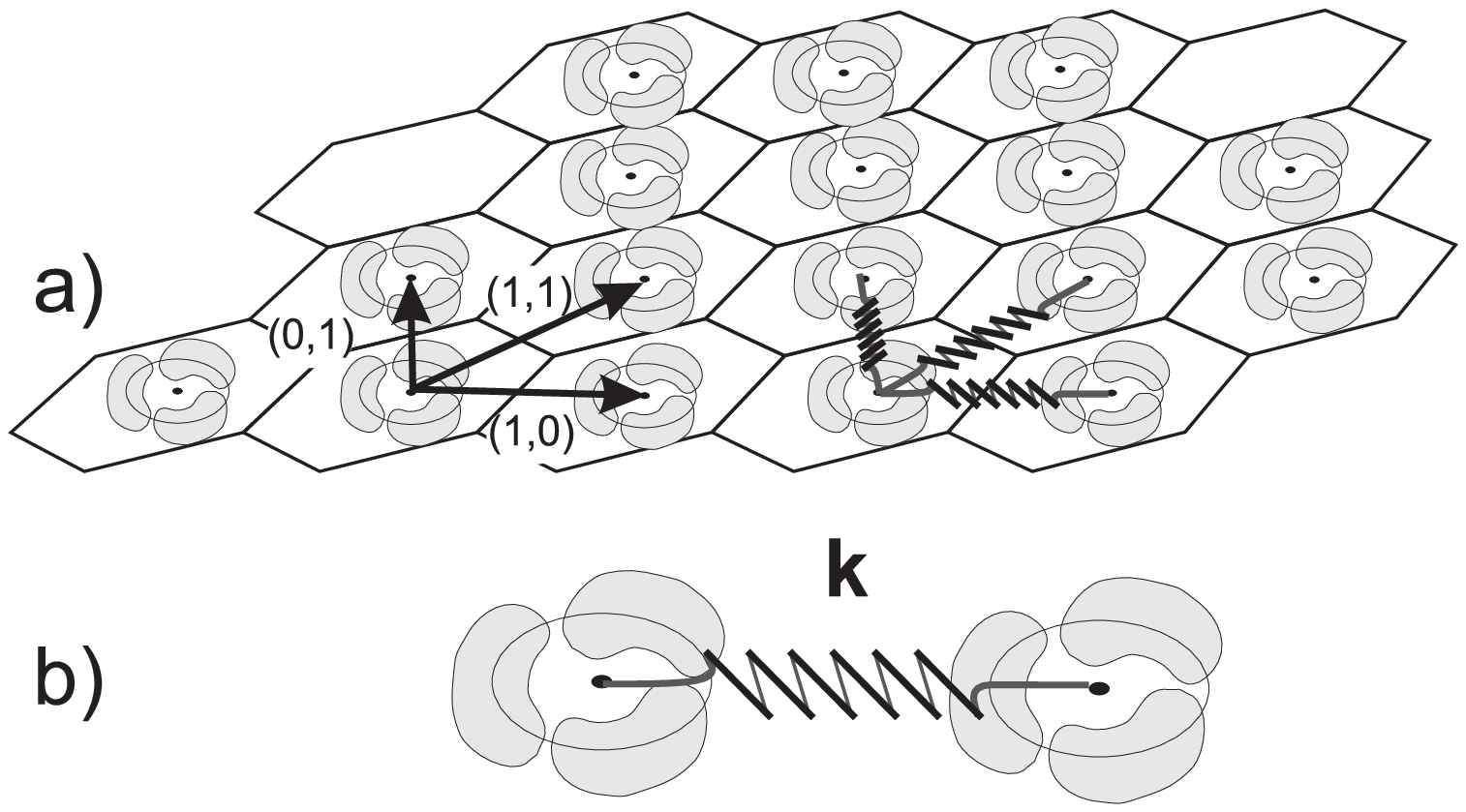}}}
\caption[]{} \label{PMSketch.eps}
\end{figure}

The high protein concentration in biological membranes might lead to
long-range protein-protein interactions, on which there have been
speculations, already some time ago \cite{Lipowsky:1995}. Recently,
interprotein motions in a carboxymyoglobin protein crystal were
reported from a molecular dynamics simulation
\cite{KurkalSiebert:2008,Meinhold:2007}. Motions in proteins occur
on various length and time scales
\cite{Frauenfelder:1991,Fenimore:2004}, and the functional behavior
of membrane proteins is likely to depend on the lipid bilayer
composition and physical properties, such as hydrophobic thickness
and elastic moduli. How the variety of inter- and intra-protein
motions, occuring over different time and length scales, interact to
result in a functioning biological system remains an open field for
those working at the interface of physics and biology. Purple
Membrane (PM) occurs naturally in the form of a two-dimensional
crystal, consisting of 75\% (wt/wt) of a single protein,
Bacteriorhodopsin (BR), that functions as a light-activated proton
pump, and 25\% various lipid species (mostly phospho- and
glyco-lipids) \cite{Haupts:1999}. BR is a proton transporting
membrane protein, formed of seven trans-membrane alpha-helices
arranged around the photosensitive retinal molecule. The protein in
the lipid matrix is organized in trimers that form a highly ordered
2d hexagonal lattice with lattice parameter $a \approx 62$~\AA, as
depicted in Fig.~\ref{PMSketch.eps}  (a).

The experiments were performed on the IN12 cold-triple-axis
spectrometer at the Institut Laue Langevin (Grenoble, France). It
allows the measurement of diffraction and inelastic scattering in
the same run without changing the set-up, which is crucial to assign
dynamical modes to structural properties and molecular components.
Correlations and motions in membranes are often well separated in
reciprocal space because of the largely different length and time
scales involved. The prominent distances in PM, such as lipid--lipid
and BR--BR monomers and trimers for instance lead to spatially well
separated signals. The same holds for the different time scales
involved from the picosecond (molecular reorientations) to the nano-
or microsecond (membrane undulations, large protein motions). The
use of oriented samples further allows to separate correlations in
the plane of the membranes, and perpendicular to the bilayers.
Dynamics between different protein trimers is expected to be
dominant where the 2d BR diffraction pattern is observed, i.e., in a
$q_{||}$ range of about 0.1~\AA$^{-1}$ to 0.6~\AA$^{-1}$.

The excitation spectrum of the 2d protein lattice was modeled
analytically, taking the protein trimers as the centers of a
primitive hexagonal lattice with lattice constant $a=62$~\AA. The
model is depicted in Fig.~\ref{PMSketch.eps}(a). The basic hexagonal
translations are marked by arrows. The interaction between the
protein trimers is contained in springs with an effective
(longitudinal) spring constant $k$ (Fig.~\ref{PMSketch.eps}(b)). The
calculated $C_l(q,\omega)$ is shown in Fig.~\ref{PMphonons.eps}(b).
The statistical average leads to a superposition of the different
phonon branches, which start and end in the hexagonal Bragg peaks
(at $\hbar\omega=0$). The absolute phonon energies can not be
determined from the model, but depend on the coupling constant $k$.
So the energy of the phonon curves in Fig.~\ref{PMphonons.eps}(b)
was scaled to match the experiment. Note that because the proteins
trimers were treated as dots with an effective mass of $M_{tr}$, the
calculation does not include any contributions from intra-protein or
intra-trimer dynamics, i.e., possible optical modes and phonons. The
longitudinal phonon spectrum
$C_l(q,\omega)=(\omega^2/q^2)S(q,\omega)$ for $q_{||}$ values
between 0.34~\AA$^{-1}$ and 0.46~\AA$^{-1}$ is shown in
Figure~\ref{PMphonons.eps}(a). The most pronounced phonon branches
from Figure~\ref{PMphonons.eps}(b) were plotted as solid lines in
part (a) for comparison. In the $q_{||}$ range between
0.34~\AA$^{-1}$ and about 0.43~\AA$^{-1}$, the experiment well
reproduces the calculated phonon curves. The agreement is less good
for $q_{||}$ values above 0.43~\AA$^{-1}$. Only the strongest phonon
branches are visible in the data, the weaker branches can most
likely not be resolved from the background.

Experiment and calculation can be compared to the MD simulations in
Ref.~[\onlinecite{KurkalSiebert:2008}], where a peak in the energy
spectrum at 1~meV was identified as a translational intermolecular
protein:protein interaction vibration in a carboxymyoglobin protein
crystal. The energy agrees well with the energy of the zone boundary
phonon in PM of 1.02~meV, as shown in Fig.~\ref{PMphonons.eps}. The
good agreement of the energy values in the two systems most likely
stems from the very high protein density in PM, which makes it
almost crystal like. The computational work thus strongly supports
the interpretation of our data as collective protein:protein
excitations. The commonly assumed interaction mechanism between
inclusions in membranes is a lipid-mediated interaction due to local
distortions of the lipid bilayer
\cite{Kralchevsky:1997,Bohinc:2003,Biscari:2002,Lague:2001,Dan:1993},
with a strong dependence on the bilayer properties, in particular
elastic properties. The PM might, however, be a special case because
there are very few lipids between neighboring BR proteins
\cite{Baudry:2001}. While the nature of the interaction  still will
 be mainly elastic, it is not likely to be purely lipid-mediated but
for the most part a direct protein--protein interaction. The
strength of the interaction can be determined from the data in
Fig.~\ref{PMphonons.eps}. The energy of the zone-boundary phonon at
the M-point of the hexagonal Brillouin zone (for instance at a
$q_{||}$ value of 0.35 \AA$^{-1}$) relates to the coupling constant
by $M_{tr} \omega^2=6k$. Because this energy is determined as $\hbar
\omega=1.02$ meV, the effective protein--protein spring constant $k$
is calculated to $k=53.49$~N/m \footnote{The molecular weight of a
BR monomer is 26.9 kDa \cite{Hunt:1997}. Because $1~{\rm kg} =
6.0221 \times 10^{26}~{\rm Dalton}$, the BR trimer thus weighs
$m_{tr}=1.34 \cdot 10^{-22}$ kg}. On the microscopic level,
displacing the BR trimer by 1 \AA\ yields a force between
neighboring trimers of 5.3~nN. There is therefore strong
protein--protein communication in PM. Using the same approach, the
spring constant for graphite for comparison is calculated to 27,000
N/m for the in-plane interaction, and 3.5 N/m for out-of-plane
interactions. The force constant that we measure in PM thus is 1-2
orders of magnitude larger than the effective van-der-Waals force
constant in graphite, but 2-3 orders of magnitude weaker than a C-C
bond.

%------------------------------------------------------------------------------------
\section{Conclusion}
Inelastic scattering experiments give access to molecular dynamics
and correlations in membranes. The experiments prove that relevant
properties and functional aspects of the bilayers can be quantified
from coherent dynamics. Future experiments will allow to study intra
and inter protein dynamics of membrane embedded proteins and how
bilayer composition and physical properties affect protein function.
Membranes with specific properties might be tailored for specific
applications and protein function of membrane embedded proteins
might be enhanced (or suppressed) by adapting the properties of the
surrounding bilayer.

{\bf Acknoledgement:} It is my pleasure to thank Christoph Ollinger,
Giovanna Fragneto, Franz Demmel, Wolfgang H\"au{\ss}ler, Karin
Schmalzl, Kathleen Wood, Dieter Strauch, who were involved in the
original work, and in particular Tim Salditt for his continuous
support and a very fruitful collaboration. The Institut
Laue-Langevin provided ample beam time to conduct the experiments.

%------------------------------------------------------------------------------------
\newpage
\bibliography{Membranes_09292008,references_dmpc2}

%------------------------------------------------------------------------------------
% Figure legends
\clearpage
\section*{Figure Legends}

\subsubsection*{Figure~\ref{cellmembrane}}
(Color online). Cartoon of a cell membrane. Courtesy of Mariana Ruiz
Villarreal
(http://commons.wikimedia.org/wiki/Image:Cell\_membrane\_detailed\_diagram.svg).

\subsubsection*{Figure~\ref{membranedynamics.eps}}
(Color online). Some elementary dynamical modes in lipid bilayers.
(a) Local modes include diffusion and vibrations, rotations and
librations (hindered rotations) of single lipid molecules. (b)
Collective excitations are coherent motions of several membrane
molecules, such as the long wavelength undulation and bending modes
of the membranes.

\subsubsection*{Figure~\ref{Collective_Function_Graph}}
(Color online). Length and time scale for some collective motions
and corresponding functional aspects. While permeability occurs on
nearest neighbor distances of lipid molecules, membrane elasticity
involves several hundreds of lipid molecules. Besides collective
intra-protein dynamics, there might also be an interaction between
different proteins in biological membranes, i.e., an inter-protein
dynamics.

\subsubsection*{Figure~\ref{dispersion_fluid_graph.eps}}
(Color online). (a) Dispersion relations at T=30$^{\circ}$C. The
solid line is a fit to Eq.~(\ref{ribotta}). (b) Dispersion relations
in the gel (19$^{\circ}$C) and in the fluid phase (22$^{\circ}$C). A
pronounced soft mode is observed at q$_0\approx$0.015\AA$^{-1}$ at
22$^{\circ}$C (dotted vertical line). Solid lines in (b) are guides
to the eye. (From \cite{RheinstaedterPRL:2006}.)

\subsubsection*{Figure~\ref{PMSketch.eps}}
(Color online). (a) BR trimers are arranged on a hexagonal lattice
of lattice constant $a\approx 62$~\AA. (b) The interaction between
the protein trimers is depicted as springs with effective spring
constant $k$. (From \cite{RheinstadterPRL:2008}.)

\begin{figure}[p] \centering
\resizebox{0.8\columnwidth}{!}{\rotatebox{0}{\includegraphics{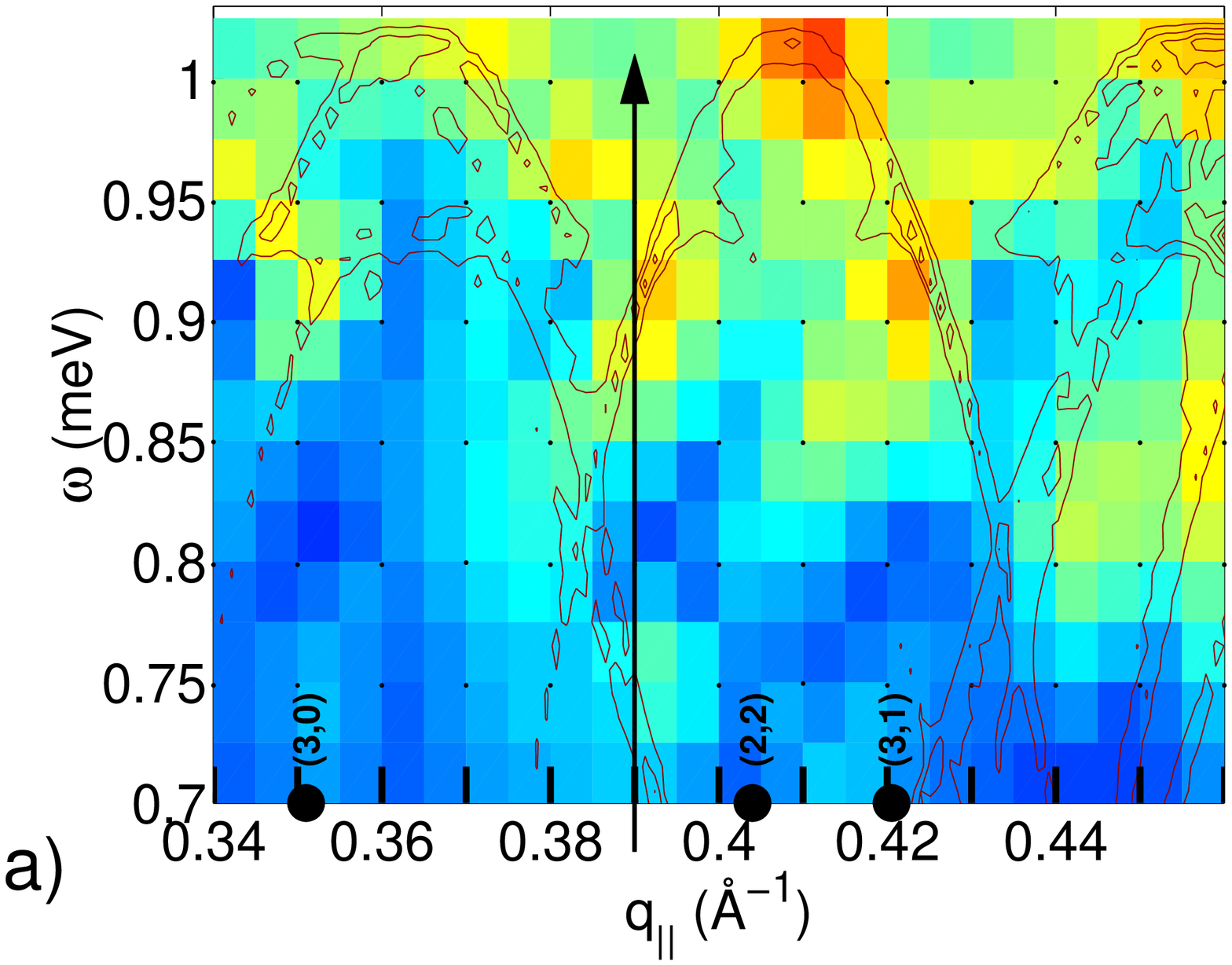}}}
\resizebox{0.8\columnwidth}{!}{\rotatebox{0}{\includegraphics{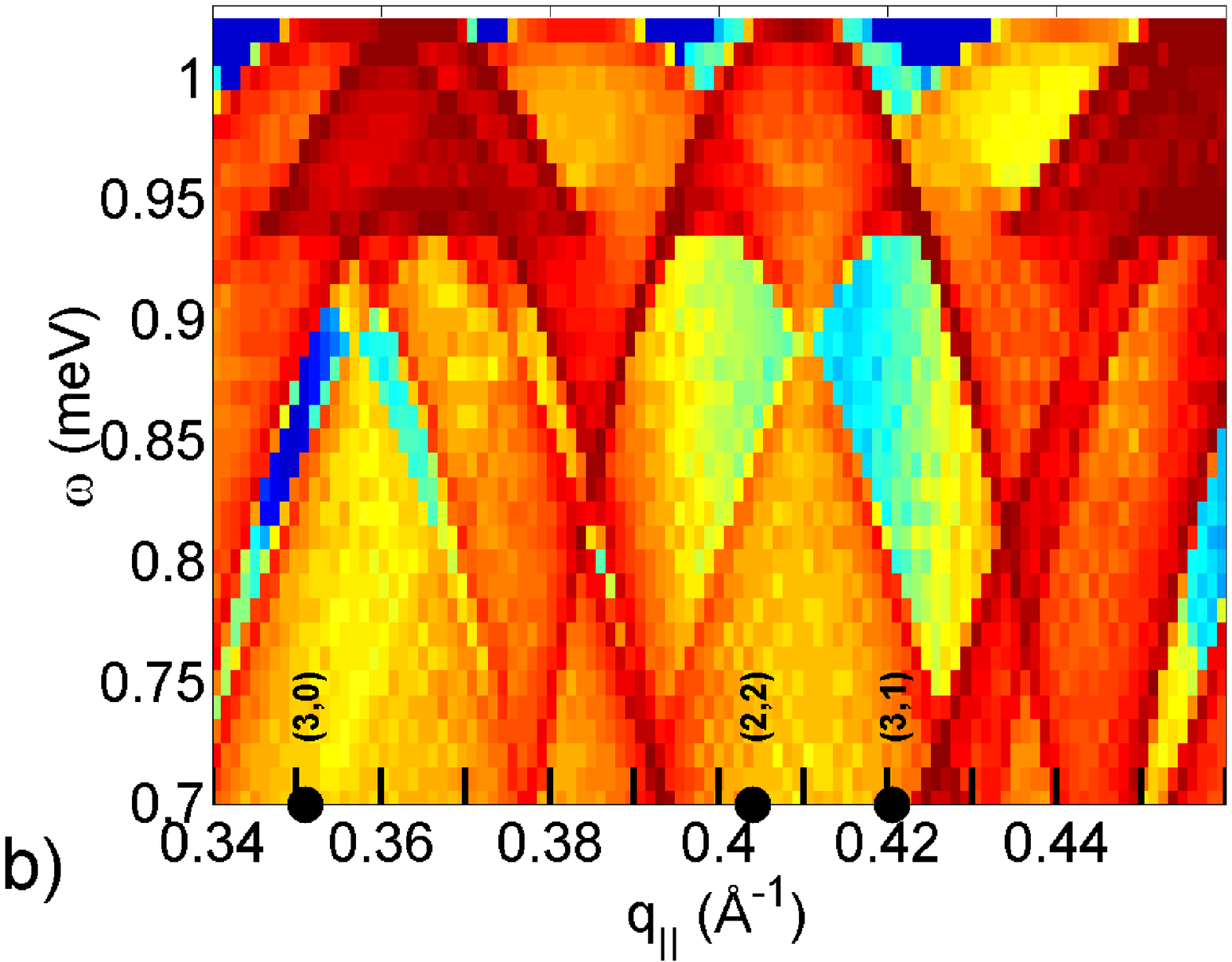}}}
\caption[]{} \label{PMphonons.eps}
\end{figure}

\subsubsection*{Figure~\ref{PMphonons.eps}}
(Color online). (a) Experimental $C_l(q,\omega)$, measured in the
the third Brillouin zone. (b) Calculated excitation spectrum
$C_l(q,\omega)$ in the range of the experimental data. The most
strongest phonon branches are also shown in part (a) as solid lines.
(From \cite{RheinstadterPRL:2008}.)

%------------------------------------------------------------------------------------
% Figures, one per page (fig_1.eps and fig_1.pdf files must be present
% in the document directory)
%\clearpage
%\section*{Figures}

\clearpage

\clearpage

\clearpage

\clearpage

\clearpage

\clearpage

%------------------------------------------------------------------------------------
\end{document}